\documentclass[a4paper,onecolumn]{IEEEtran}

\usepackage{psfrag}

\usepackage{amsmath,amssymb,amstext}

\usepackage{pstricks}
\usepackage{epsfig}
\usepackage{pst-grad} 
\usepackage{pst-plot} 

\usepackage{setspace} 		
\newcommand{\nn}{\nonumber \\}
\newcommand{\lucin}[1]{#1}
\newcommand{\lucout}[1]{}
\DeclareMathOperator{\D}{d\!}
 \DeclareMathOperator{\ee}{e}

\begin{document}

\title{L2 Orthogonal Space Time Code for Continuous Phase Modulation}

\author{Matthias Hesse, J\'er\^ome Lebrun and Luc Deneire\thanks{The work of Matthias Hesse is supported by the EU by a Marie-Curie Fellowship (EST-SIGNAL program : http://est-signal.i3s.unice.fr) under contract No
MEST-CT-2005-021175.}\thanks{The authors are with Lab. I3S, CNRS, University of Nice, Sophia Antipolis, France; E-Mail:\tt \{hesse,lebrun,deneire\}@i3s.unice.fr}}

\markboth{L2 Orthogonal Space Time Code for Continuous Phase Modulation}{}

\maketitle
\thispagestyle{empty}
\begin{abstract}
  To combine the high power efficiency of Continuous Phase Modulation
  (CPM) with either high spectral efficiency or enhanced performance
  in low Signal to Noise conditions, some authors have proposed to introduce
  CPM in a MIMO frame, by using Space Time Codes (STC).  In this
  paper, we address the code design problem of Space Time Block Codes
  combined with CPM and introduce a new design criterion based on
  $L^2$ orthogonality.  This $L^2$ orthogonality condition, with the
  help of simplifying assumption, leads, in the 2x2 case, to a new
  family of codes.  These codes generalize the Wang and Xia code,
  which was based on pointwise orthogonality.  Simulations indicate
  that the new codes achieve full diversity and a slightly better
  coding gain.  Moreover, one of the codes can be interpreted as two
  antennas fed by two conventional CPMs using the same data but with
  different alphabet sets.  Inspection of these alphabet sets lead
  also to a simple explanation of the (small) spectrum broadening of
  Space Time Coded CPM.

\end{abstract}
\section{Introduction}
\label{sec:intro}
Since the pioneer work of Alamouti \cite{Alam98} and Tarokh
\cite{Taro99a}, Space Time Coding  has been a fast growing field
of research where numerous coding schemes have been introduced.
Several years later Zhang and Fitz \cite{Zhan00,Zhan03} were the first
to apply the idea of STC to continuous phase modulation (CPM) by
constructing trellis codes. In \cite{Zaji07} Zaji\'c and St\"uber
derived conditions for partial response STC-CPM to get full diversity
and optimal coding gain.  A STC for noncoherent detection based on
diagonal blocks was introduced by Silvester et al. \cite{Silv06}.

The first orthogonal STC for CPM for full and partial response was
developed by Wang and Xia \cite{Wang04,Wang05}. The scope of this
paper is also the design of an orthogonal STC for CPM. But unlike
Wang-Xia aprroach \cite{Wang05} which starts from a QAM orthogonal
Space-Time Code (e.g. Alamouti's scheme \cite{Alam98}) and modify it
to achieve continuous phases for the transmitted signals, we show here
that a more general $L^2$ condition is sufficient to ensure fast
maximum likelihood decoding with full diversity.

\begin{figure}
 \centering
  \includegraphics[width=4in]{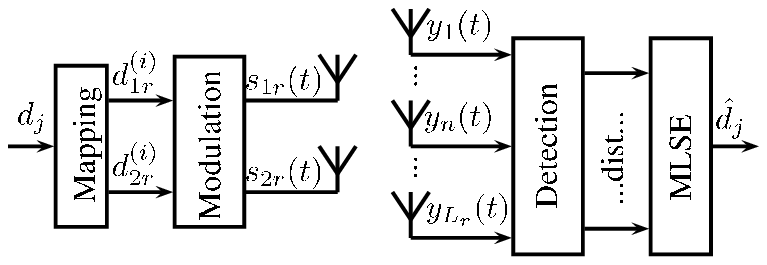}
  \caption{\small Structure of a MIMO Tx/Rx system }
  \label{fig:block}
\end{figure}

In the considered system model (Fig.\ref{fig:block}), the data
sequence $d_j$ is defined over the signal constellation set
\begin{equation}
  \Omega_{d}=\{ -M+1,-M+3,\ldots,M-3,M-1\}
\end{equation}
for an alphabet with $\log_2M$ bits. To obtain the structure for a
\lucin{Space Time} Block Code (\lucin{ST}BC) this sequence is mapped
to data matrices $\mathbf D^{(i)}$ with elements $d_{mr}^{(i)}$, where
$m$ denotes the transmitting antenna, $r$ the time slot into a block and
$(i)$ a parameter for partial response CPM. The data matrices are then
used to modulate the sending matrix
\begin{equation}
  \label{eq:symmatrix}
  {\bf S}(t) = \begin{bmatrix}
    s_{11}(t) & s_{12}(t) \\
    s_{21}(t) & s_{22}(t)
  \end{bmatrix}\text{.}
\end{equation}
Each element is defined for $(2l+r-1)T\leq t\leq(2l+r)T$ as
\begin{equation}
  s_{mr}(t) = \sqrt{\frac{E_s}{T}}\ee^{j2\pi\phi_{mr}(t)} 
\end{equation}
where $E_s $ is the symbol energy and $T$ the symbol time. The phase
$\phi_{mr}(t)$ is defined in the conventional CPM manner \cite{Ande86}
with an additional  correction factor $c_{mr}(t)$ and is
therewith given by
\begin{equation}
  \phi_{mr}(t)=\theta_m(2l+r) + 
  h\!\!\!\!\!\!\!\!\!\sum\limits_{i=2l+1+r-\gamma}^{2l+r}
  \!\!\!\!\!\!\!\!\!d_{mr}^{(i)}q(t-(i-1)T)+c_{mr}(t)
  \label{eq:phase}
\end{equation}
where $h=2m_0/p$ with $m_0$ and $p$ relative primes is called the
modulation index. The phase smoothing function $q(t)$ has to be a
continuous function with $q(t)=0$ for $t\leq0$ and $q(t)=1/2$ for
$t\geq \gamma T$.

The memory length $\gamma$ determines the length of $q(t)$ and affects
the spectral compactness. For large $\gamma$ we obtain a compact
spectrum but also a higher number of possible phase states which
increases the decoding effort. For full response CPM, we have
$\gamma=1$ and for partial response systems $\gamma>1$.

 The choice of the correction factor $c_{mr}(t)$ in Eq.
(\ref{eq:phase}) is along with the mapping of $d_j$ to $\mathbf
D^{(i)}$, the key element in the design of our coding scheme. It will
be detailed in Section \ref{sec:Cond}. We then define $\theta_m(2l+r)$
in a most general way
\begin{align}
  \theta_m(2l+3)&=\theta_m(2l+2)+\xi(2l+2) \nn
  &=\theta_m(2l+1)+\xi(2l+1)+\xi(2l+2).
  \label{eq:theta}
\end{align}
The function $\xi(2l+r)$ will be fully defined from the contribution
$c_{mr}(t)$ to the phase memory $\theta_m(2l+r)$. For conventional CPM
system, $c_{mr}(t)=0$ and we have
$\xi(2l+1)=\frac{h}{2}d_{2l+1-\gamma}$.

 The channel coefficients $\alpha_{mn}$ are assumed to be
Rayleigh distributed and independent. Each coefficient $\alpha_{mn}$
characterizes the fading between the $\mathrm{m^{th}}$ transmit (Tx) antenna and
the $\mathrm{n^{th}}$ receive (Rx) antenna where $n=1,2,\ldots,L_r$.
Furthermore, the received signals
\begin{equation}
  y_n(t)=\alpha_{mn}s_{mr}(t)+n(t)	
\end{equation}
are corrupted by a complex additive white Gaussian noise $n(t)$ with
variance $1/2$ per dimension.

 At the receiver, the detection is done on each of the $L_r$
received signals separately. Therefore, in general, each code block
$\mathbf S(t)$ has to be detected by block. E.g. for a 2x2 block,
estimating the symbols $\hat d_j$ implies \lucout{an effort}
\lucin{computational complexity} proportional to $M^2$. Now, this
\lucout{effort}\lucin{complexity} can be reduced to $2M$ by
\lucout{introducing some properties}\lucin{introducing an
  orthogonality property as well as simplifying assumptions} on the
code.

Criteria for such ST\lucin{B}C are given in Section \ref{sec:Cond}. In
Section \ref{sec:OSTC}, the criteria are used to construct
OST\lucin{B}C for CPM. In Section \ref{sec:simul} we test the designed
code and compare it with the STC from Wang and Xia \cite{Wang05}.
Finally, some conclusions are drawn in Section \ref{sec:concl}.

\section{Design Criteria}
\label{sec:Cond}

The purpose of the design is to achieve full diversity and a fast
maximum likelihood decoding while maintaining the continuity of the
signal phases.  \lucin{This section shows how the need to perform fast
  ML decoding leads to the $L^2$ orthogonality condition as well as to
  simplifying assumptions, which can be combined with the continuity
  conditions.} For convenience we only consider one Rx antenna and drop the 
  index $n$ in $\alpha_{mn}$.

\lucin{\subsection{Fast Maximum Likelihood Decoding}} 

Commonly, due to the trellis structure of CPM, the Viterbi algorithm is used 
to perform the ML demodulation . On block $l$ each state in the trellis has $M^2$ incoming
branches and $M^2$ outgoing branches with a distance

\begin{align}
  D_l&= \int\limits_{2lT}^{(2l+1)T} \Big|
      y(t)-\sum\limits_{m=1}^2 \alpha_{m}s_{m1}(t) \Big| ^2\D
    t+\int\limits_{(2l+1)T}^{(2l+2)T} \Big|
      y(t)-\sum\limits_{m=1}^2 \alpha_{m}s_{m2}(t) \Big|^2\D t.
  \label{eq:MLext}
\end{align}
The number of branches results from the blockwise decoding and the
 correlation between the sent symbols $s_{1r}(t)$ and $s_{2r}(t)$. 
 A way to reduce the number of branches  is to
structurally decorrelate the signals sent by the two transmitting
antennas, i.e. to put to zero the inter-antenna correlation
\begin{multline}
  \alpha_{2}\alpha_{1}^*\!\!\!\!\!\!\int\limits_{2lT}^{(2l+1)T}
  \!\!\!\!\!\! s_{21}(t)s_{11}^*(t)\D t+\alpha_{1}\alpha_{2}^*\!
  \!\!\!\!\!\!\! \int\limits_{2lT}^{(2l+1)T} \!\!\!\!\!\!\!\!
  s_{11}(t)s_{21}^*(t)\D t + \alpha_{2}\alpha_{1}^*
  \!\!\!\!\!\!\!\int\limits_{(2l+1)T}^{(2l+2)T}\! \!\!\!\!\!\!
  s_{22}(t)s_{12}^*(t)\D t+\alpha_{1}\alpha_{2}^*\!
  \!\!\!\!\!\!\int\limits_{(2l+1)T}^{(2l+2)T}\! \!\!\!\!\!\!
  s_{12}(t)s_{22}^*(t)\D t = 0.
  \label{eq:cross0}
\end{multline}

Pointwise orthogonality as defined in \cite{Wang05} is therefore a
sufficient condition but not necessary. A less restrictive $L^2$
orthogonality is also sufficient. From Eq. (\ref{eq:cross0}), the distance 
given in Eq.  (\ref{eq:MLext}) can then be simplified to

\begin{align}
  D_l&=\int\limits_{2lT}^{(2l+1)T}\!\!\!\!\!\!\!
    f_{11}(t)+f_{21}(t)-|y(t)|^2 \D t+\int\limits_{(2l+1)T}^{(2l+2)T}\!\!\!\!\!\!\!
    f_{12}(t)+f_{22}(t) -|y(t)|^2 \D t 
\end{align}
with $f_{mr}(t)=|y(t)-\alpha_m s_{mr}(t)|^2$. When each $s_{mr}(t)$
depends only on $d_{2l+1}$ or $d_{2l+2}$ the branches can be split and 
calculated separately  for  $d_{2l+1}$ and $d_{2l+2}$. The
complexity of the ML decision is reduced to $2M$. The complexity for
detecting two symbols is thus reduced from $pM^{\gamma+1}$ to
$pM^{\gamma}$. The STC introduced by Wang and Xia \cite{Wang05}
didn't take full advantage of the orthogonal design since $s_{mr}(t)$
was depending on both $d_{2l+1}$ and $d_{2l+2}$. The gain they
obtained in \cite{Wang05} was then relying on other properties of CPM,
e.g. some restrictions put on $q(t)$ and $p$. These
restrictions may also be applied to our design code, which would lead
to additional complexity reduction. However, this is not in the scope
of this paper and is be the subject of another upcoming paper.

\subsection{Orthogonality Condition}

In this section we show how $L^2$ orthogonality for CPM, i.e.
$\|\mathbf S(t)\|^2_{L_2}=\int_{2lT}^{(2l+2)T}\mathbf S(t)\mathbf S^H(t)\D t=2\mathbf I$, can
be obtained. As such, the correlation between the two transmitting
antennas per coding block is cancelled if

\begin{align}
  \int\limits_{2lT}^{(2l+2)T}s_{1r}(t)s_{2r}^*(t)\D
  t&=\int\limits_{2lT}^{(2l+1)T}s_{11}(t)s_{21}^*(t)\D t +  \int\limits_{(2l+1)T}^{(2l+2)T}s_{12}(t)s_{22}^*(t)\D t=0\text{.}
\end{align}
 Replacing $s_{mr}(t)$ by the corresponding CPM symbols from Eq.
(\ref{eq:phase}), we get\small
\begin{align}
     \!\int\limits_{2lT} ^{(2l+1)T}\!\!\!\!\!\!\!  \exp\!\Big\{j2\pi\big[\theta_1(2l+1)+   h\!\!\!\!\!\!\!\sum\limits_{i=2l+2-\gamma}^{2l+1}\!\!\!\!\!\!\! d_{1,1}^{(i)} q(t-(i-1)T)+c_{1,1}(t)-
       \theta_2(2l+1)-h\!\!\!\!\!\!\sum\limits_{i=2l+3-\gamma}^{2l+2}\!\!\!\!\!\! d_{2,1}^{(i)}q(t-(i-1)T)-c_{2,1}(t)\big]\Big\}\D t+ \nn
            \!\int\limits_{(2l+1)T} ^{(2l+2)T} \!\!\!\!\!\!\!\exp\!\Big\{j2\pi\big[\theta_1(2l+2)+h\!\!\!\!\!\!\!\sum\limits_{i=2l+3-\gamma}^{2l+2}\!\!\!\!\!\!\!d_{1,2}^{(i)}q(t-(i-1)T)\!+\!c_{1,2}(t)-
     \theta_2(2l+2)-h\!\!\!\!\!\!\sum\limits_{i=2l+2-\gamma}^{2l+1}  \!\!\!\!\!\!d_{2,2}^{(i+1)}q(t-iT) -c_{2,2}(t) \big]\D t\Big\} = 0.
\end{align}\normalsize
The phase memory $\theta_m(2l+r)$ is independent of time and has not
to be considered for integration. Using Eq. (\ref{eq:theta}) to
replace phase memory $\theta_m(2l+2)$ of the second time slot, we
obtain\small
\begin{align}
  \int\limits_{2lT} ^{(2l+1)T}  \!\!\!\!\!\! \exp\Big\{j2\pi\big[h\!\!\!\!\!\!\sum\limits_{i=2l+2-\gamma} ^{2l+1} \!\!\!\!\!\!  d_{1,1}^{(i)}q(t-(i-1)T) +c_{1,1}(t)-
  h\!\!\!\!\!\!\sum\limits_{i=2l+2-\gamma}^{2l+1} \!\!\!\!\!\!d_{2,1}^{(i)}q(t-(i-1)T)-c_{2,1}(t)\big]\Big\}\D t+&\nn 
  \exp\Big\{ j2\pi\big[\xi_1(2l+1)-\xi_2(2l+1)\big]\Big\}\cdot
  \int\limits_{2lT}^{(2l+1)T}\!\!\!\!\!\!\! \exp \Big\{j2\pi\big[h\!\!\!\!\!\!\sum\limits_{i=2l+2-\gamma}^{2l+1}\!\!\!\!\!\!d_{1,2}^{(i+1)}   q(t-(i-1)T)+c_{1,2}(t+T)-&\nn
   h\!\!\!\!\!\!\sum\limits_{i=2l+2-\gamma}^{2l+1}  \!\!\!\!\!\!d_{2,2}^{(i+1)}q(t-(i-1)T) -c_{2,2}(t+T) \big]\Big\}\D t &=  0.
  \label{eq:orthLong}
\end{align}\normalsize

\lucin{\subsection{Simplifying assumptions}} To simplify this
expression, we \lucin{factor} \lucout{split} Eq. (\ref{eq:orthLong}) into a time
independent and a time dependent part. For merging the two integrals
to one time dependent part, we have to map $d_{m2}^{(i)}$ to
$d_{m1}^{(i)}$ and $c_{mr}(t)$ to a different
$c_{m\lucin{'}r\lucin{'}}(t)$. Consequently, for the data symbols
$d_{mr}^{(i)}$ there exist three possible ways of mapping:
\begin{itemize}
\item {\em crosswise mapping} with $d_{1,1}^{(i)}=d_{2,2}^{(i)}$ and
  $d_{1,2}^{(i)}=d_{2,1}^{(i)}$;
\item {\em repetitive mapping} with   $d_{1,1}^{(i)}=d_{1,2}^{(i)}$ and
  $d_{2,1}^{(i)}=d_{2,2}^{(i)}$;
\item {\em parallel mapping} with $d_{1,1}^{(i)}=d_{2,1}^{(i)}$ and
  $d_{1,2}^{(i)}=d_{2,2}^{(i)}$ .
\end{itemize}
The same approach can be applied to $c_{mr}(t)$:
\begin{itemize}
\item {\em crosswise mapping} with $c_{11}(t)=-c_{22}(t-T)$ and
  $c_{12}(t)=-c_{21}(t-T)$;
\item {\em repetitive mapping} with $c_{11}(t)=c_{12}(t-T)$ and $  c_{21}(t) =c_{22}(t-T)$;
\item {\em parallel mapping} with $c_{11}(t)=c_{21}(t)$ and
  $c_{12}(t)=c_{22}(t)$.
\end{itemize}

 For each combination of mappings, Eq. (\ref{eq:orthLong}) is now
\lucout{ in two products}\lucin{the product of two factors}, one
containing the integral and \lucin{the other} a time independent part.
To fulfill Eq. (\ref{eq:orthLong}) it is sufficient if one factor is
zero, namely $ 1+\ee^{j2\pi\left[\xi_1(2l+1)-\xi_2(2l+1)\right]} =0$,
i.e. if
\begin{equation}
   k+\frac{1}{2}=\xi_1(2l+1)-\xi_2(2l+1)
  \label{eq:orthCond}
\end{equation}
with $k\in \mathbb N$. We thus get a very simple condition which only
depends on $\xi_m(2l+1)$.

\subsection{Continuity of Phase}

In this section we determine the functions $\xi_m(2l+1)$ to ensure the
phase continuity.

Precisely,  the phase of the CPM symbols has to be equal at all
intersections of symbols. For an arbitrary block $l$, it means that
$\phi_{m1}((2l+1)T) =\phi_{m2}((2l+1)T)$. Using Eq. (\ref{eq:phase}),
it results in
\begin{align}
  \xi_m(2l+1)&= h \!\!\!\!\!\!\!\!\!\sum\limits_{i=2l+2-\gamma}^{2l+1}
  \!\!\!\!\!\!\!\! d_{m,1}^{(i)}q((2l+2-i)T)+c_{m,1}((2l+1)T) -h \!\!\!\!\!\!\!\sum\limits_{i=2l+3-\gamma}^{2l+2}
  \!\!\!\!\!\!\! d_{m2}^{(i)}q((2l+2-i)T)-c_{m2}((2l+1)T)\text{.}
  \label{eq:xi1org}
\end{align}
 For the second intersection at $(2l+2)T$, since
$\phi_{m2}((2l+2)T)=\phi_{m1}((2l+2)T)$, we get
\begin{align}
  \xi_m(2l+2)&=h \!\!\!\!\!\!\sum\limits_{i=2l+3-\gamma}^{2l+2}
  \!\!\!\!\!\!\! d_{m2}^{(i)}q((2l+3-i)T)+c_{m2}((2l+2)T)-h  \!\!\!\!\!\!\!\!\!\!\sum\limits_{i=2(l+1)+2-\gamma}^{2(l+1)+1}
  \!\!\!\!\!\!\!\!\!\!\!
  d_{m,1}^{(i)}q((2l+3-i)T)-c_{m,1}((2l+2)T)]\text{.}
  \label{eq:xi2org}
\end{align}

 Now, by choosing one of the mappings detailed in Section
\ref{sec:OSTC}, these equations can be greatly simplified. Hence, we
have all the tools to construct our code.

\section{Orthogonal Space Time Codes}
\label{sec:OSTC}

In this section we will have a closer look at two codes constructed
under the afore-mentioned conditions.

\subsection{Existing Code}

As a first example, we will give an alternative construction of the
code given by Wang and Xia in \cite{Wang05}.  \lucin{Indeed, the
  pointwise orthogonality condition used by Wang and Xia is a special
  case of the $L^2$ orthogonality condition, hence, their ST-code can
  be obtained within our framework.} \lucout{ Namely, as expected, this
  ST-code can be obtained within our framework. It is easily verified
  to be orthogonal in $L^2$ sense but this is not surprising since
  Wang and Xia constructed their code from pointwise orthogonality
  (Alamouti scheme) and this is a special case of $L^2$
  orthogonality.}

For the first antenna Wang and Xia use a conventional CPM with
$d_{1r}^{(i)}=d_i$ for $i=2l+r+1-\gamma,2l+r+2-\gamma,\ldots ,2l+r$
and $c_{1r}(t)=0 $. The symbols of the second antenna are mapped
{ \em crosswise} to the first $d_{21}^{(i)}=-d_{i+1}$ for
$i=2l+2-\gamma,2l+3-\gamma,\ldots ,2l+1$ and $d_{22}^{(i-1)}=-d_{i-1}$
for $i=2l+3-\gamma,2l+4-\gamma,\ldots ,2l+2$. Using this cross mapping
makes it difficult to compute $\xi_m(2l+1)$ since the CPM typical
order of the data symbols is mixed. Wang and Xia circumvent this by
introducing another correction factor for the second antenna
\begin{equation}
  c_{2r}(t)\!=\!\!\sum\limits_{i=0}^{\gamma-1}(h(
  {d_{2l+1-i}+d_{2l+2-i}})+1 ) q_0(t-(2l+r-1-i)T)
\end{equation}
By first computing $\xi_m(2l+1)$ with Eq.  (\ref{eq:xi1}) and then Eq.
(\ref{eq:orthCond}), we get the $L^2$ orthogonality of the
Wang-Xia-STC.

\subsection{Parallel Code}

To get a simpler correction factor as in \cite{Wang05}, we
designed a new code based on the {\em parallel} structure which permits
to maintain the conventional CPM mapping for both antennas. Hence we
choose the following mapping: $d_{m1}^{(i)}=d_{m2}^{(i-1)}=d_i$ for
$i=2l+r+1-\gamma,2l+r+2-\gamma,\ldots ,2l+r$. Then, Eq.
(\ref{eq:xi1org}) and (\ref{eq:xi2org}) can be simplified into
\begin{align}
  \xi_m(2l+1)&=\frac{h}{2}d_{2l+2-\gamma}+c_{m1}((2l+1)T)-c_{m2}((2l+1)T) \label{eq:xi1}\\
  \xi_m(2l+2)&=\frac{h}{2}d_{2l+3-\gamma}+c_{m2}((2l+2)T)-c_{m1}((2l+2)T)\text{.}
\end{align}
 With this simplified functions, the orthogonality condition
only depends on the start and end values of $c_{mr}(t)$, i.e.
\begin{align}
  k+\frac{1}{2}&=c_{11}((2l+1)T)-c_{12}((2l+1)T)-c_{21}((2l+1)T)+c_{22}((2l+1)T).
  \label{eq:condC}
\end{align}
To merge the two integrals in Eq. (\ref{eq:orthLong}) not
only the mapping of $d_{mr}^{(i)}$ is necessary but also an equality
between different $c_{mr}(t)$. From the three possible mappings, we
choose the {\em repeat mapping} because of the possibility to set
$c_{mr}(t)$ to zero for one antenna. Hence we are able to send a
conventional CPM signal on one antenna and a modified one on the
second.  Using Eq.  (\ref{eq:condC}) and the equalities for the
mapping, we can formulate the following condition
\begin{equation}
  k+\frac{1}{2}=c_{12}(2lT)-c_{12}((2l+1)T)-c_{22}(2lT)+c_{22}((2l+1)T).
  \label{eq:cR}
\end{equation}


 With $c_{11}(t)=c_{12}(t)=0$, we can take for
$c_{21}(t)=c_{22}(t)$ any continuous function which is zero at $t=0$
and $1/2$ at $t=T$.  Another possibility is to choose the correction
factor of the second antenna with a structure similar to CPM
modulation, i.e.
\begin{equation}
  c_{2r}(t)=\sum\limits_{i=2l+1-\gamma}^{2l+1}q(t-(i-1)T)
\end{equation}
for ${(2l+r-1)T\leq t\leq(2l+r)T}$.  With this approach, the
correction factors can be included in a classical CPM modulation with
constant offset of $1/h$. This offset may also be expressed as a
modified alphabet for the second antenna
\begin{equation}
  \Omega_{d_2}=\{ -M+1+\frac{1}{h},-M+3+\frac{1}{h},\ldots,M-3+\frac{1}{h},M-1+\frac{1}{h}\}
  \label{eq:omega2}
\end{equation}

Consequently, this $L^2$-orthogonal design may be seen as two
conventional CPM designs with different alphabet sets $\Omega_d$ and
$\Omega_{d2}$ for each antenna. However, in this method, the constant
offset to the phase may cause a shift in frequency. But as shown by
our simulations in the next section, this shift is quite moderate.

\section{Simulations}
\label{sec:simul}

\begin{figure}
	\centering
  \includegraphics[width=3.5in]{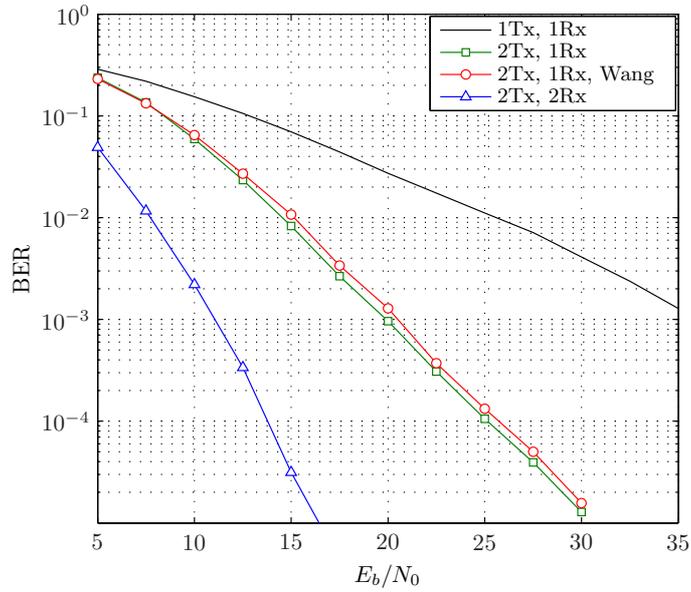}
  \caption{\small Simulated BER for different numbers of Tx and Rx antennas of
    the proposed STC and of the Wang-Xia-STC }
  \label{fig:BER}
\end{figure}

In this section we verify the proposed algorithm by simulations.
Therefore a STC-2REC-CPM-sender with two transmitting antennas has
been implemented in MATLAB. For the signal of the first antenna we use
conventional Gray-coded CPM with a modulation index $h=1/2$, the
length of the phase response function $\gamma=2$ and an alphabet size
of $M=8$. The signal of the second antenna is modulated by a CPM with
the same parameters but a different alphabet $\Omega_{d_2}$,
corresponding to Eq. (\ref{eq:omega2}).

The channel used is a frequency flat Rayleigh fading model with
additive white Gaussian noise. The fading coefficients $\alpha_{mn}$
are constant for the duration of a code block
(\lucout{fast}\lucin{block} fading) and known at receiver (coherent
detection). The received signal $y_n(t)$ is demodulated by two
filterbanks with $pM^2$ filters, which are used to calculate the
correlation between the \lucout{truly} received and \lucout{possible
  received}\lucin{candidate} signals. Due to the orthogonality of the
antennas each filterbank is independently applied to the corresponding
time slot $k$ of the block code. The correlation is used as metric for
the Viterbi algorithm (VA) which has $pM$ states and $M$ paths
leaving each state.  \lucout{For}\lucin{In our} simulation\lucin{,}
the VA is truncated to a path memory of 10 code blocks, which means
that we get a decoding delay of $2\cdot 10T$.

\lucout{Figure \ref{fig:BER} shows the results of our simulations. The
  continuous line without marker is the result of a system without
  diversity (1 transmitting (Tx) and 1 receiving (Rx) antenna) and is
  used as reference. The curve marked with circles results from the
  coding introduced by Wang and Xia \cite{Wang05} using 2 Tx and 1 Rx
  antenna.  The plots marked with buttons and triangles we obtained by
  the coding introduced in this paper with 2 Tx and 1 or 2 Rx
  antennas, respectively.

  As expected, the use of multiple independent channels yields an
  improved BER behavior. The reference system with one channel (1 Tx
  and 1 Rx antenna) has a slope of slightly more than 10dB per decade
  for high EbNo. For 2 Tx antennas and 1 Rx antenna we obtain a slope
  of around 5 dB per decade for both the STC of Wang and Xia and our
  approach. A doubling of the slope while doubling transmitter
  antennas at the same time, corresponds to full diversity. The small
  difference between the proposed STC and the one of Wang and Xia may
  be explained by our simpler design. Finally, by doubling the number
  of receiving antennas we reach a steepness of a bit less than 3 dB
  per decade, which approximately corresponds to full diversity. Due
  to these simulation results we can reasonably assume that the
  proposed code achieves full diversity.  }

\begin{figure}
  \includegraphics[width=3.5in]{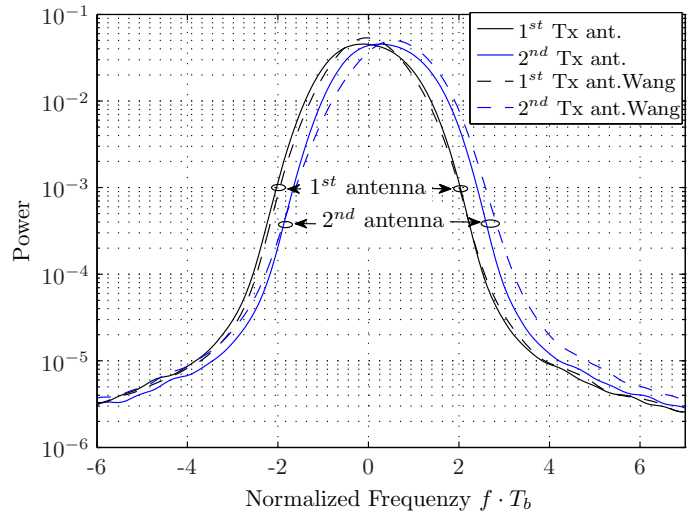}
  \centering
  \caption{\small Simulated psd for each Tx antenna of the proposed STC
    (continuous line) and the Wang-Xia-STC (dashed line)}
  \label{fig:Spec}
\end{figure}

\lucin{From the simulation results given in Figure \ref{fig:BER}, we
  can reasonably assume that the proposed code achieves full
  diversity.  Indeed, the curves for the 2x1 and 2x2 systems
  respectively show a slope of 2 and 4.  Moreover, the curve of the
  2x1 systems follows the same slope as the ST code proposed by Wang
  and Xia \cite{Wang05}, which was proved to have full diversity.
  Note also that the new code provides a slightly better performance.}

A main reason of using CPM for STC is the spectral efficiency. Figure
\ref{fig:Spec} show the simulated power spectral density (psd) for
both Tx antennas of the proposed ST code (continuous line) and the ST
code proposed by Wang and Xia \cite{Wang05}. The first antenna of our
approach uses a conventional CPM signal and hence shows an equal psd.
The spectrum of the second antenna is shifted due to adding an offset
$c_{mr}(t)$ with a non zero mean. Minimizing the difference between
the two spectra by shifting one, result in a phase difference of
$0.375$ measured in normalized frequency $f\cdot T_d$, where
$T_d=T/\log_2(M)$ is the bit symbol length. The first antenna of the
Wang-Xia-algorithm has almost the same psd while the spectrum of the
second antenna is shifted by approximately $0.56 f\cdot T_d$. This
means that the OSTC by Wang and Xia requires a slightly larger
bandwidth than our OSTC.

\section{Conclusion}
In applications where the power efficiency is crucial, combination of
Continuous Phase Modulation and Space Time Coding has the potential to
provide high spectral efficiency, thanks to spatial diversity. To
address this power efficiency, ST code design for CPM has to
ensure both low complexity decoding and full diversity.  To fulfill
these requirements, we have proposed a new $L^2$ orthogonality
condition.  We have shown that this condition is sufficient to achieve
low complexity ML decoding and leads, with the help of simplifying
assumption to a simple code.  Moreover, simulations indicate that the
code most probably achieves full diversity.  Further work will be
concentrated on the design of other codes based on $L^2$ orthogonality as 
in the meanwhile, we have been able to obtain the design of full diversity, 
full rate $L^2$ orthogonal codes for 3 antennas \cite{Hess08}.
\label{sec:concl}
\small \bibliographystyle{IEEEbib} \bibliography{IEEEabrv,bib}

\end{document}